\documentclass[12pt]{article}

\usepackage{scicite}
\usepackage{times}

\usepackage{graphicx} 

\usepackage[articletitle=true,style=science,maxbibnames=99]{biblatex}
\bibliography{main.bib}

\usepackage{url}
\usepackage{comment}

\usepackage{setspace}

\newenvironment{sciabstract}{%
\begin{quote} \bf}
{\end{quote}}

\title{Strategic COVID-19 vaccine distribution can simultaneously elevate social utility and equity}

\author
{Lin Chen,$^{1\dagger}$ Fengli Xu,$^{2\dagger}$ Zhenyu Han,$^{4}$ Kun Tang,$^{5}$ \\ Pan Hui,$^{1\ast}$ James Evans,$^{2,3\ast}$ Yong Li$^{4\ast}$\\
\\
\normalsize{$^{1}$Department of Computer Science and Engineering,}\\ 
\normalsize{The Hong Kong University of Science and Technology, Hong Kong SAR, P. R. China}\\
\normalsize{$^{2}$Knowledge Lab \& Department of Sociology, University of Chicago, Chicago, IL, USA}\\
\normalsize{$^{3}$Santa Fe Institute, Santa Fe, NM, USA}\\
\normalsize{$^{4}$Beijing National Research Center for Information Science and Technology (BNRist),}\\
\normalsize{Department of Electronic Engineering, Tsinghua University, Beijing, P. R. China}\\
\normalsize{$^{5}$Vanke School of Public Health, Tsinghua University, Beijing, P. R. China}\\
\\
\normalsize{$\dagger$Lin Chen and Fengli Xu contribute equally to this work.}\\
\normalsize{$^\ast$To whom correspondence should be addressed;} \\
\normalsize{E-mail: panhui@cse.ust.hk, jevans@uchicago.edu, liyong07@tsinghua.edu.cn.}
}

\date{}

\begin{document} 

\baselineskip24pt

\maketitle


\begin{sciabstract}

Abstract: Balancing social utility and equity in distributing limited vaccines represents a critical policy concern for protecting against the prolonged COVID-19 pandemic. What is the nature of the trade-off between maximizing collective welfare and minimizing disparities between more and less privileged communities?
To evaluate vaccination strategies, we propose a novel epidemic model that explicitly accounts for both demographic and mobility differences among communities and their association with heterogeneous COVID-19 risks, then calibrate it with large-scale data.
Using this model, we find that social utility and equity can be simultaneously improved when vaccine access is prioritized for the most disadvantaged communities, which holds even when such communities manifest considerable vaccine reluctance.
Nevertheless, equity among distinct demographic features are in tension due to their complex correlation in society.
We design two behavior-and-demography-aware indices, \textit{community risk} and \textit{societal harm}, which capture the risks communities face and those they impose on society from not being vaccinated, to inform the design of comprehensive vaccine distribution strategies. 
Our study provides a framework for uniting utility and equity-based considerations in vaccine distribution, and sheds light on how to balance multiple ethical values in complex settings for epidemic control.
\end{sciabstract}


\noindent \textbf{One-Sentence Summary:} 
Collective welfare and equity among communities can be improved together with behavior-and-demography-aware vaccine distribution strategies.
\\

\noindent \textbf{Main Text:}
Sweeping the globe since its outbreak, the COVID-19 pandemic continues to impose harm and loss on human societies everywhere~\cite{whocoviddashboard}. This has revealed and intensified disparities in the health risks faced by different communities because disadvantaged groups are less resilient to adversity~\cite{wang2020combating,gray2020covid,van2020covid}. 
In the fight against the pandemic, vaccines are regarded as the most critical medical resources, but still face prominent shortages in many countries and communities~\cite{padma2021covid,maxmen2021covid}. As a result, vaccine prioritization has becomes a critical policy task in every public health system~\cite{prioritize_eu,prioritize_us,jhuinterim} with well-considered strategies to balance social utility and equity.
Utility and equity represent the most visible metrics considered by health authorities and organizations worldwide~\cite{mcclung2020advisory,national2020framework,jhuinterim,world2020sage}, undergirded by the contrasting ethical values of utilitarianism and egalitarianism.
As Jeremy Bentham put it, ``the well-being of a portion of individuals'' can sometimes be sacrificed to achieve ``the greatest happiness of the greatest number''~\cite{bentham1781introduction}. 
Previous research has identified trade-offs between social utility and equity in the distribution of healthcare resources ranging from disease screening to treatment~\cite{sassi2001equity,ottersen2008distribution, world2000world, lindholm1996epidemiological, dolan1998measurement}.
In COVID-19 vaccine distribution, the most recent studies have focused on the trade-off between minimizing years of life and number of lives lost~\cite{goldstein2021vaccinating,castro2021prioritizing,dushoff2021transmission,buckner2021dynamic,bubar2021model}, both of which reflect social utility-oriented values and neglect disparities across population. 
In light of this gap, we aim to reveal a critical relationship between social utility and equity in COVID-19 vaccine distribution with implications for designing vaccine prioritization. 

Examining the social utility and equity of vaccine distribution strategies requires an epidemiological model adequate to capture the uneven risks faced by different communities~\cite{buckner2021dynamic}, \emph{e.g.}, older persons and those with greater mobility are associated with higher COVID-19 risk~\cite{gray2020covid,o2021age,carrion2021neighborhood,weill2020social}. 
However, standard epidemiological models, \emph{e.g.}, SIR~\cite{bailey1975mathematical,nunn2006infectious} and SEIR~\cite{li1995global,mwalili2020seir}, are built on the assumption of homogeneous population mixing, which prohibits them to capture heterogeneity in the spread of coronavirus.
Some recent work~\cite{chang2021mobility,gozzi2021estimating,lai2020effect} has aimed to augment standard epidemiological models with empirical mobility data, but these neglect inherent vulnerability differences embedded in demographic profiles~\cite{o2021age}.
Here, we propose a novel epidemiological model that simultaneously captures heterogeneity in the mobility patterns and demographic profiles of different communities. 
Calibrated with large-scale mobility and census data covering more than 75 million residents in the United States, our model automatically tunes the dynamics of coronavirus spread within each community based on its demographic profile and traces dispersion among communities with time-varying empirical mobility networks. 
Our model can predict the number of daily deaths in each metro statistical area (MSA) and reconstruct its uneven distribution among communities with accuracy, enabling the evaluation of equity among communities defined by different demographics, \emph{i.e.,} percentage of older adults, average household income, and percentage of essential workers, which have garnered considerable attention throughout the pandemic ~\cite{van2020covid,o2021age,fitzpatrick2021optimizing,chang2021mobility,carrion2021neighborhood,mena2021socioeconomic,weill2020social}. Accordingly, we examine three vaccine distribution strategies that prioritize communities based on vulnerabilities defined by these three demographic features. 
We find that social utility and equity can be simultaneously improved when prioritizing the most disadvantaged communities in each demographic dimension for vaccine access. Such a result holds even under scenarios where low-income communities show considerable vaccine hesitancy, which contrasts with the conventional view of inevitable trade-offs~\cite{ottersen2008distribution,world2000world,lindholm1996epidemiological,dolan1998measurement}.
Nevertheless, elevating the equity across one demographic dimension can degrade it across others, suggesting that demographic features alone are insufficient to guide vaccine distribution.
We propose two demography-and-behavior-aware indices, \textit{community risk} and \textit{societal harm}, designed to measure the effect of prioritizing each community for vaccination to reduce (1) its own mortality risks and (2) the mortality risks it imposes on society as a whole.
Based on these two indices, we design a framework for vaccine prioritization that simultaneously improves social utility and equity in all dimensions across scenarios of different vaccination rates and timing.
By providing a general framework to tease out utilitarian and egalitarian values in COVID-19 vaccine distribution, our findings carry broad implications for the design of vaccination strategies.

\section*{Behavior and demography informed epidemic modelling}

To capture the heterogeneity in health risk faced by different communities~\cite{mwalili2020seir,chang2021mobility,gozzi2021estimating}, we propose a novel epidemic model that principally integrates two important factors in the spread of coronavirus (Fig. 1a), \emph{i.e.}, demographic profiles and mobility behaviors. 
Specifically, demographic profile is found to be significantly correlated with the fatality rate of SARS-CoV-2 infection~\cite{o2021age,goldstein2020demographic}, while mobility behavior determines the likelihood of exposure to coronavirus~\cite{chang2021mobility,lai2020effect,mena2021socioeconomic,glaeser2020jue}. 
Therefore, the proposed \underline{B}ehavior and \underline{D}emography informed epidemic model (BD) divides the studied population based on the minimum geographical units defined by the United States Census Bureau, known as census block groups (CBGs)~\cite{cbgdef}, and maintains a local susceptible-exposed-infectious-recovered (SEIR) model for each of them to characterize the dynamics of intra-CBG epidemic spread, where the infection-fatality rate (IFR) is adjusted based on the demographic profile and age-specified risks estimated in previous medical research~\cite{o2021age}. 
To capture inter-CBG transmission resulting from urban mobility, the proposed BD model constructs a bipartite network linking CBGs and points of interest (POIs) with time-varying edges to track hourly movements extracted from the SafeGraph dataset~\cite{safegraphweekly}, where edge weights reflect temporal mobility intensity. 
New infections occur in POIs and CBGs with differing epidemic parameters capturing environmental characteristics, and infected populations travel to other communities proportional to their extracted movements (Materials and Methods, M1).

We evaluate the proposed BD model on nine large metro statistical areas (MSAs) in the United States covering over $75$ million people, and compare it with two baseline models: a standard SEIR model and a meta-population model that only considers heterogeneous mobility among communities~\cite{chang2021mobility}. 
Results show that the proposed BD model consistently produces more accurate estimations of daily deaths in each MSA with growth patterns ranging from sub-linear to exponential (Fig. 1b).
Specifically, the BD model outperforms the SEIR and meta-population models by reducing up to $80\%$ and $77\%$ of the normalized root mean square error, respectively (Fig. 1c, Fig. S1). 
The proposed BD model also reveals higher mortality risks faced by communities with higher percentages of older adults, lower household income, and essential workers, consistent with real-world observations~\cite{van2020covid,o2021age,chang2021mobility,carrion2021neighborhood,mena2021socioeconomic,weill2020social,davies2021community,gozzi2021estimating} (Fig. 1d). 
In contrast, the meta-population model predicts communities with higher percentages of older adults to face unreasonably lower mortality risk, likely due to its inability to model demographic profiles. 
Further, the SEIR model cannot capture heterogeneous risks in different communities due to its assumption of homogeneous population mixing.

We examine the correlations between demographic profiles and mobility behavior to explain heterogeneity across communities predicted by the meta-population model and our proposed BD model (Fig. 1e). 
We find that the percentage of older adults negatively correlates with per capita mobility ($r=-0.29$), and thus neglect of demography-specific mortality risk will lead to inaccurate estimation of risk for different age groups. 
By contrast, our proposed BD model finds differences in mobility behaviors outweighed by the change in IFRs due to age structure and predicts higher mortality risk in communities with higher percentages of older adults, consistent with previous research~\cite{o2021age,davies2021community}. 
Moreover, communities with lower average household incomes and higher percentages of essential workers are associated with higher levels of mobility, likely following from limitations in their ability to significantly reduce mobility during the pandemic~\cite{gozzi2021estimating,weill2020social,chang2021mobility}.
Therefore, both the meta-population model and proposed BD model reproduce higher risk associated with low-income communities, while the risk associated with essential worker percentages remains complicated due to the joint effect of demographic and mobility profiles, \emph{e.g.}, essential workers generally have higher mobility but younger demographic profiles. 
In view of this, considering the joint effect of both mobility behaviors and demographic profiles should enable improved prediction of the heterogeneous risks facing different communities. 
By incorporating both into the epidemic model and utilizing large-scale real-world mobility data for calibration, the proposed BD model is effective in generating accurate daily predictions and capturing heterogeneous risks faced by distinctive communities and provides a framework from which we can analyze the impact of differing vaccine distribution strategies on social utility and equity.

\section*{Consequences of alternative vaccine distribution strategies}

Social utility and equity represent the two most important concerns considered by public health policy makers in the COVID-19 pandemic~\cite{jhuinterim,mcclung2020advisory,national2020framework,world2020sage}. These account for the collective welfare of the entire society and disparities among individual communities. 
As a critical policy concern during the COVID-19 pandemic, discussion of vaccine distribution strategies have centered on the trade-off between social utility and equity~\cite{jama_equity}, but this has been inadequately evaluated or supported with empirical data.
With the proposed BD model, we aim to reveal mechanisms behind social utility and equity in vaccine distribution with large-scale empirical data.
In vaccine distribution, we quantify social utility as a reduction in the overall fatality rate, and we quantify equity as a reduction in the Gini coefficient of fatality rates among communities (Materials and Methods, M4).
Based on previous analyses that reveal heterogeneous health risks faced by populations with different ages, income levels and occupations (Fig. 2d), we focus on three dimensions of equity among communities, \emph{i.e.}, equity among age groups, income groups and occupational groups. 

Prioritizing the least advantaged populations is acknowledged as a fundamental value in healthcare resource allocation~\cite{persad2009principles,doi:10.1056/NEJMsb2005114,emanuel2020ethical}.  Based on this, we construct three vaccine distribution strategies that prioritize the most disadvantaged communities defined on three dimensions of demographic profiles, \emph{i.e.}, percentage of older adults (\textit{Prioritize by age}), average household income (\textit{Prioritize by income}) and percentage of essential workers (\textit{Prioritize by occupation}).  
As a baseline for comparison, we also construct a \textit{Homogeneous} strategy, which provides vaccine access to each community with uniform probability~\cite{macnaughton2009untangling}.
Experiments show strategies that prioritize the worst-off communities drastically improve the equity defined on the corresponding demographic dimension (Fig. \ref{fig2}a). 
Specifically, equity defined on age, income and essential worker percentage can be improved by $28.0\%$, $43.9\%$ and $45.6\%$ compared with the \textit{Homogeneous} baseline, respectively.
Moreover, in most cases all three strategies also achieve improvement in social utility compared with the baseline (Fig. \ref{fig2}a), suggesting that it is highly probable to simultaneously improve overall social utility and equity defined on a specific demographic dimension by prioritizing the most disadvantaged communities within that dimension. 
This sharply contradicts the conventional view of an inevitable trade-off between social utility and equity~\cite{ottersen2008distribution,world2000world,lindholm1996epidemiological,dolan1998measurement}. 
Detailed analysis reveals that prioritizing communities under greater risks consistently results in larger improvements for both social utility and equity, which further highlights the effectiveness achieved by prioritizing the worst-off communities (Table S2).

We further incorporate the impact of vaccine hesitancy~\cite{khubchandani2021covid,dror2020vaccine,loomba2021measuring}, the phenomenon wherein people mistrust and refuse to take vaccines despite access.
According to the empirically observed association between low income and low vaccine acceptance~\cite{khubchandani2021covid}, we construct five scenarios to investigate the impact of this phenomena on a vaccine distribution strategy focused on the most disadvantaged communities.
In the first two scenarios, \textit{Real-1} and  \textit{Real-2}, we set the vaccine acceptance rates for different communities according to a population sample-based national assessment \cite{khubchandani2021covid}. 
In \textit{Real-2}, we further combine estimated vaccine acceptance rates with empirical vaccination data from the U.S. CDC~\cite{vac_demo}, to calculate vaccine accessibility in each community.
In three hypothetical scenarios, we set vaccine acceptance rates from bottom to top income groups at  $0.6,0.7,0.8,0.9,1$ (\textit{Hypo-1}), $0.2,0.4,0.6,0.8,1$ (\textit{Hypo-2}) and $0.1,0.3,0.5,0.7,1$ (\textit{Hypo-3}), sequentially reflecting larger differences across income groups and allowing us to explore more extreme disparities in vaccine hesitancy than we currently observe (Materials and Methods, M2).
In general, the improvement in social utility diminishes as differences in vaccine acceptance rates grow larger (Fig. \ref{fig2}c).
Among experimented prioritization strategies, \textit{Prioritize by income} is most sensitive to changes in vaccine acceptance rates.
Nevertheless, its improvement to social utility does not vanish until the vaccine acceptance rate in the bottom income group drops to $1/5$ of that in the top group, as in \textit{Hypo-2}, explained by vastly disproportionate risks facing different income groups in the pandemic. This hypothetical hesitancy is far larger than what we observe from data.
As would be expected, in all five scenarios, prioritizing the most disadvantaged communities consistently and significantly improves equity.
Following the robust guideline to prioritize the most disadvantaged, social utility and equity can both be improved even if the most disadvantaged groups manifest the most vaccine hesitancy. It demonstrates the outsized protective impact that would accrue to society from far greater investments in vaccination outreach, education and incentive for vaccination for our most disadvantaged communities.

For each strategy, we also calculate its impact on equity defined along other demographic dimensions. We find it is difficult to achieve a comprehensive improvement in all dimensions by simply prioritizing the worst-off communities (Fig. \ref{fig2}b). 
To reveal the relationship between disadvantages along different demographic dimensions, we analyze correlations among the percentage of older adults, average household income, and percentage of essential workers for each CBG (Materials and Methods, M3). 
We observe a positive correlation between percentage of older adults and average household income ($r=+0.14$), indicating that populations with older demographics tend to have higher income. 
Moreover, the percentage of essential workers negatively correlates with the percentage of older adults ($r=-0.2$), but positively correlates with average household income ($r=+0.28$), indicating that populations with a larger proportion of essential workers tend to be younger with higher household incomes (Fig. \ref{fig2}d). 
These correlations suggest the mismatch of disadvantaged populations in different demographic dimensions, which results in conflicts between equities that cannot be settled based on demographic features alone.
This suggests the need to explore more essential mechanisms underlying demographic features and health that can forecast vaccination outcomes and guide vaccine distribution.

\section*{Proposing indices for estimating vaccine prioritization outcomes}

To inform the design of vaccine distribution, it is critical to accurately estimate the outcomes that would result from prioritizing certain communities for vaccine distribution.
Specifically, for optimal design we must be able to estimate changes in overall social utility and equity when vaccinating each community. 
Policy designers typically rely either on a single demographic feature~\cite{schmidt2021equitable} or indicators computed solely based on demographic data, such as the social vulnerability index (SVI) designed by the U.S. Center for Disease Control~\cite{svi2018}. 
Nevertheless, Fig.~2 shows that such approaches will likely degrade equity along certain dimensions, due primarily to complex associations between distinct demographic profiles, infection-fatality risks, and per capita mobility (Fig. S2). 
Therefore, we design two novel vaccine outcome indices, \textit{community risk} and \textit{societal harm} that capture these two important mechanisms.
In order to evaluate changes in equity when vaccinating a community, we design a \textit{community risk} index as the expected mortality rate, calculated as the product of estimated contact frequency due to average community movement and infection-fatality risk associated with community demographic profiles. 
In order to evaluate changes in risks to society when vaccinating a community, we design a \textit{societal harm} index as the expected number of deaths caused by infection in that community and the secondary infections they impose on persons from other communities (Materials and Methods, M5). Thus, it captures the number of lives saved in the whole population by vaccinating certain communities, providing a proxy for social utility.

Our proposed \textit{community risk} and \textit{societal harm} indices capture different characteristics of communities that could potentially result in trade-offs between social utility and equity. 
For example, older people with lower mobility face greater mortality risk once they are infected, but they are less likely to spread the disease to other people compared to young people with high mobility. This is manifested by their high \textit{community risk} and low \textit{societal harm} indices, respectively. 

We perform regression analysis to examine the power of these indices for estimating the outcomes associated with alternative vaccination distribution strategies. 
Specifically, we generate numerous vaccine distribution instances, each of which vaccinates a set of randomly selected communities covering $2\%$ of the total population. 
We obtain the impact on social utility and equity for each vaccine distribution through simulation, and perform ordinary least-squares (OLS) regression to estimate changes with demographic features and proposed indices (Materials and Methods, M5). 
Results show that: (1) Demographic features alone only explain on average $28.0\%$ of the variance in fatality rate reduction, but the incorporation of our \textit{societal harm} index significantly raises that value to $62\%$ (Fig. \ref{fig3}b). 
(2) Demographic features alone only explain on average $53.0\%$, $37.3\%$ and $29.9\%$ of the variances in equity defined on age, income and essential worker percentage, respectively, 
but incorporation of the \textit{community risk} index raises those values to $66.4\%$, $49.6\%$ and $43.8\%$, respectively (Fig. \ref{fig3}c, detailed regression results presented in Table S4-S12).
The indices of \textit{societal harm} and \textit{community risk} significantly improve estimation of changes in social utility and equity under any community prioritization scheme.
These two indices also shed light on the simultaneous improvement of social utility and equity with vaccination. 
Specifically, we discover a positive correlation between \textit{community risk} and \textit{societal harm} (Fig. \ref{fig3}d), which indicates a non-negligible overlap between communities experiencing large \textit{community risk} and those imposing large \textit{societal harm}.
Therefore, if a vaccine distribution strategy succeeds in targeting such overlapping communities, it can simultaneously achieve improvement in both social utility and equity.

\section*{Informing the design of vaccine distribution strategies}

Based on our proposed indices of vaccination outcomes, we design a flexible framework to generate well-rounded vaccine distribution strategies (\emph{i.e.}, a \textit{Comprehensive} strategy) that can improve both social utility and equity in all demographic dimensions.
Our framework integrates \textit{community risk}, \textit{societal harm} and  demographic profiles with learned weights to generate a comprehensive index of vaccine priority for each community, then distributes vaccines by communities accordingly (Materials and Methods, M6).
Besides the \textit{Homogeneous} baseline and three strategies examined in Fig. 2, we also construct two additional strategies for comparison.
First, an \textit{SVI-informed} strategy is designed to prioritize vaccines to communities according to the social vulnerability index (SVI) released by the U.S. CDC~\cite{svi2018} , which is recommended for use in vaccine prioritization~\cite{schmidt2021equitable,svi_michigan}.
Second, to further justify the necessity of \textit{community risk} and \textit{societal harm}, we construct a \textit{Comprehensive-ablation} strategy that utilizes demographic features without the indices (Materials and Methods, M6). 

Results show that the \textit{Comprehensive} strategy yielded by our framework achieves greater improvements in both social utility and all dimensions of equity, in comparison to the \textit{Homogeneous} baseline (Fig. \ref{fig4}a-c).  
In contrast, all the other prioritization strategies, whether based on demographic features (percentage of older adults, average household income, percentage of essential workers) or indicators calculated solely from demographic data (SVI), degrade either social utility or certain dimensions of equity, and thus fail to strike an optimal balance.
Specifically, although \textit{Comprehensive-ablation} is informed by the same set of demographic features as \textit{Comprehensive}, it is still unable to guarantee improvements in all health outcome measures because it does not incorporate the impact of mobility. 
Therefore, demographic features are inadequate to guide the design of vaccine distribution strategies, but our proposed vaccination outcome indices, \textit{community risk} and \textit{societal harm}, complete the framework and fulfill this task.

To examine the generalizability of this framework, we further construct two series of experimental scenarios that reflect different levels of vaccine supply and epidemic intensity.
To estimate the overall performance of a vaccination strategy, we take the sum of relative changes in social utility and three dimensions of equity, which approximates calculation of an L1-norm for a vector, but taking sign into account.
In the first series of scenarios, we vary the vaccination rate from $5\%$ to $56\%$ of the total population, reflecting different vaccine supply levels (Fig. \ref{fig4}d). 
Specifically, the vaccination rate of $56\%$ reflects vaccination progress in the U.S. by October, 2021.
In this scenario, we construct an additional strategy, $Real\mbox{-}world$, which distributes vaccines proportional to the real-world distribution estimated by the U.S. CDC~\cite{vac_demo}.
In the second series of scenarios, we change the timing of vaccination by up to $10$ days to reflect different levels of pervasive epidemic spreading (Fig. \ref{fig4}e). 
Experiments of finer-grained changes in vaccination rate and timing can be found in the supplementary materials (Fig. S3-S17).
In all variations, our framework successfully yields comprehensive strategies that simultaneously elevate the social utility and three dimensions of equity.
We note that the $Real\mbox{-}world$ strategy is highly limited in overall performance, suggesting there remains substantial space to improve real-world vaccine distribution strategies even with high vaccination rates.
Projected improvement is more prominent with lower vaccination rates, which is reasonable because with increased vaccine supply, overlapping vaccinated populations will expand and eventually eliminate the differences between prioritization strategies.
Nevertheless, this trend highlights that in the face of a greater vaccine shortage, more attention should be paid to coordinate elevation of overall welfare and the mitigation of health disparities.
In sum, our experiments demonstrate that our vaccine distribution framework is generalizable across different metro areas, epidemic burdens and vaccination timings.

\section*{Discussion}

Coordination among multiple ethical values is of central concern when limited, critical healthcare resources such as vaccines must be apportioned to people in the face of profound health crises.
Different from the traditional view of unavoidable trade-offs between social utility and equity, our \underline{B}ehavior and \underline{D}emography informed epidemic model (BD) reveals that prioritizing the most disadvantaged communities for COVID-19 vaccine access can simultaneously improve social utility and equity. This outcome is driven by underlying community heterogeneity in both mobility behavior and demographic profile.
We resolve the tension of equity across different demographic features by designing two comprehensive indices, \textit{community risk} and \textit{societal harm}, to estimate vaccination outcomes.
The effectiveness of both indices shows the necessity to jointly consider both demographic and behavioral heterogeneity in epidemic modelling and policy design.

The vaccine distribution framework we propose can provide clear guidance to policy-makers.
Currently, prioritization of vaccines is usually based on a rigid stratification of age or occupation~\cite{schmidt2021equitable,fitzpatrick2021optimizing,matrajt2021vaccine}, and set for the state or country as a whole.
In contrast, our framework enables the design of flexible distribution strategies that are aware of the joint effects from mobility behaviors and demographic profiles, which can be tailored to local conditions.  
Moreover, our framework possesses the following two benefits.
First, our method provides as meso-scale policy guidance by achieving a balance between effectiveness and ease of implementation.
With awareness of heterogeneous risks faced by different CBGs, we can maximize the benefits to society with limited vaccination dosages.  
Meanwhile, because vaccine priorities are determined on the CBG level, people within the same CBG are not discriminated against, providing local administrations with greater flexibility in the actual vaccine roll-out. 
Second, our method is privacy-preserving.
For both demographic features and mobility records, we only make use of aggregate data on the CBG level, without revealing any individual information, presenting only a minimal invasion of personal privacy.
Therefore, our framework is not only theoretically informative, but also instructive for real-world practice.

More broadly, equitable access to immunization is viewed by many as a critical part of \textit{the right to health}, a fundamental human right endowing every person with the ability to pursue and claim their highest attainable health status~\cite{backman2008health,PILLAY20082005}.
Although it has been listed among the six principles of the Global Vaccine Action Plan (GVAP) since 2011~\cite{2013B5}, there is still a long way to go before we reach this ultimate goal.
Our research provides insights to settle concerns regarding ethical values in the global health crisis with experiments on large, real-world data. 

Our study has several limitations.
First, because medical studies on the etiology and pathogenesis of coronavirus are still ongoing, 
we only consider the widely-acknowledged heterogeneity in IFR associated with age.
Second, we solely focus on vaccine distribution within a country, but as the COVID-19 pandemic is a global public health emergency, it is of equal necessity to quantitatively study how to coordinate social utility and equity at the international level~\cite{world2020sage,emanuel2020ethical}.
Third, our study focuses only on the demand-side, and future studies should also include supply-side issues, as vaccine distribution is a complex process involving multiple players and systems beyond the healthcare system alone~\cite{Golan2021}.

Nevertheless, our study provides powerful guidance for vaccine prioritization even under circumstances of extreme vaccine hesitancy, recommending far greater societal investments in vaccination outreach, education and incentives for disadvantaged and undervaccinated communities than have hitherto been explored. Vaccinating those worst off represents the best step toward societal protection.   

\nocite{o2021age}
\nocite{chang2021mobility}
\nocite{goldstein2020demographic}
\nocite{nytdata}
\nocite{ew-unitedway}
\nocite{worldbanklabor}
\nocite{world2020sage}
\nocite{khubchandani2021covid}
\nocite{vac_demo}
\nocite{illsley1987measurement}
\nocite{leclerc1990differential}
\nocite{berndt2003measuring}
\nocite{dixon1987bootstrapping}
\nocite{delaware-list}
\nocite{glaeser2020jue}
\nocite{sanchez2020jobs}
\nocite{dudley2020disparities}
\nocite{household_size}
\nocite{svi2018}
\nocite{tzeng2011multiple}


\printbibliography[title={References and Notes}]

@article{gray2020covid,
  title={COVID-19 and the other pandemic: populations made vulnerable by systemic inequity},
  author={Gray, Darrell M and Anyane-Yeboa, Adjoa and Balzora, Sophie and Issaka, Rachel B and May, Folasade P},
  journal={Nature Reviews Gastroenterology \& Hepatology},
  volume={17},
  number={9},
  pages={520--522},
  year={2020},
  publisher={Nature Publishing Group}
}

@article{van2020covid,
  title={COVID-19 exacerbating inequalities in the US},
  author={Van Dorn, Aaron and Cooney, Rebecca E and Sabin, Miriam L},
  journal={Lancet (London, England)},
  volume={395},
  number={10232},
  pages={1243},
  year={2020},
  publisher={Elsevier}
}

@article{o2021age,
  title={Age-specific mortality and immunity patterns of SARS-CoV-2},
  author={O’Driscoll, Megan and Dos Santos, Gabriel Ribeiro and Wang, Lin and Cummings, Derek AT and Azman, Andrew S and Paireau, Juliette and Fontanet, Arnaud and Cauchemez, Simon and Salje, Henrik},
  journal={Nature},
  volume={590},
  number={7844},
  pages={140--145},
  year={2021},
  publisher={Nature Publishing Group}
}

@article{chang2021mobility,
  title={Mobility network models of COVID-19 explain inequities and inform reopening},
  author={Chang, Serina and Pierson, Emma and Koh, Pang Wei and Gerardin, Jaline and Redbird, Beth and Grusky, David and Leskovec, Jure},
  journal={Nature},
  volume={589},
  number={7840},
  pages={82--87},
  year={2021},
  publisher={Nature Publishing Group}
}

@article{carrion2021neighborhood,
  title={Neighborhood-level disparities and subway utilization during the COVID-19 pandemic in New York City},
  author={Carri{\'o}n, Daniel and Colicino, Elena and Pedretti, Nicolo Foppa and Arfer, Kodi B and Rush, Johnathan and DeFelice, Nicholas and Just, Allan C},
  journal={Nature Communications},
  volume={12},
  number={1},
  pages={1--10},
  year={2021},
  publisher={Nature Publishing Group}
}

@article{mena2021socioeconomic,
  title={Socioeconomic status determines COVID-19 incidence and related mortality in Santiago, Chile},
  author={Mena, Gonzalo E and Martinez, Pamela P and Mahmud, Ayesha S and Marquet, Pablo A and Buckee, Caroline O and Santillana, Mauricio},
  journal={Science},
  volume={372},
  number={6545},
  year={2021},
  publisher={American Association for the Advancement of Science}
}

@article{weill2020social,
  title={Social distancing responses to COVID-19 emergency declarations strongly differentiated by income},
  author={Weill, Joakim A and Stigler, Matthieu and Deschenes, Olivier and Springborn, Michael R},
  journal={Proceedings of the National Academy of Sciences},
  volume={117},
  number={33},
  pages={19658--19660},
  year={2020},
  publisher={National Acad Sciences}
}

@article{mcclung2020advisory,
  title={The Advisory Committee on Immunization Practices’ ethical principles for allocating initial supplies of COVID-19 vaccine—United States, 2020},
  author={McClung, Nancy and Chamberland, Mary and Kinlaw, Kathy and Matthew, Dayna Bowen and Wallace, Megan and Bell, Beth P and Lee, Grace M and Talbot, H Keipp and Romero, Jos{\'e} R and Oliver, Sara E and others},
  journal={Morbidity and Mortality Weekly Report},
  volume={69},
  number={47},
  pages={1782},
  year={2020},
  publisher={Centers for Disease Control and Prevention}
}

@article{national2020framework,
%  title={Framework for equitable allocation of COVID-19 vaccine},
%  author={National Academies of Sciences, Engineering, and Medicine and others},
%  year={2020},
%  publisher={National Academies Press}
%}

@techreport{world2020sage,
  title={WHO SAGE roadmap for prioritizing the use of COVID-19 vaccines in the context of limited supply: an approach to inform planning and subsequent recommendations based upon epidemiologic setting and vaccine supply scenarios, 13 November 2020},
  %author={World Health Organization},
  year={2020},
  institution={World Health Organization}
}

@book{world2000world,
  title={The world health report 2000: health systems: improving performance},
  %organization={World Health Organization},
  %author={World~Health~Organization},
  year={2000},
  publisher={World Health Organization}
}

@article{lindholm1996epidemiological,
  title={An epidemiological approach towards measuring the trade-off between equity and efficiency in health policy},
  author={Lindholm, Lars and Rosen, M{\aa}ns and Emmelin, Maria},
  journal={Health Policy},
  volume={35},
  number={3},
  pages={205--216},
  year={1996},
  publisher={Elsevier}
}

@book{bailey1975mathematical,
  title={The mathematical theory of infectious diseases and its applications},
  author={Bailey, Norman TJ and others},
  number={2nd ediition},
  year={1975},
  publisher={Charles Griffin \& Company Ltd 5a Crendon Street, High Wycombe, Bucks HP13 6LE.}
}

@book{nunn2006infectious,
  title={Infectious diseases in primates: behavior, ecology and evolution},
  author={Nunn, Charles and Altizer, Sonia and Altizer, Sonia M},
  year={2006},
  publisher={Oxford University Press}
}

@article{li1995global,
  title={Global stability for the SEIR model in epidemiology},
  author={Li, Michael Y and Muldowney, James S},
  journal={Mathematical biosciences},
  volume={125},
  number={2},
  pages={155--164},
  year={1995},
  publisher={Elsevier}
}

@article{mwalili2020seir,
  title={SEIR model for COVID-19 dynamics incorporating the environment and social distancing},
  author={Mwalili, Samuel and Kimathi, Mark and Ojiambo, Viona and Gathungu, Duncan and Mbogo, Rachel},
  journal={BMC Research Notes},
  volume={13},
  number={1},
  pages={1--5},
  year={2020},
  publisher={BioMed Central}
}

@article{goldstein2021vaccinating,
  title={Vaccinating the oldest against COVID-19 saves both the most lives and most years of life},
  author={Goldstein, Joshua R and Cassidy, Thomas and Wachter, Kenneth W},
  journal={Proceedings of the National Academy of Sciences},
  volume={118},
  number={11},
  year={2021},
  publisher={National Acad Sciences}
}

@article{castro2021prioritizing,
  title={Prioritizing COVID-19 vaccination by age},
  author={Castro, Marcia C and Singer, Burton},
  journal={Proceedings of the National Academy of Sciences},
  volume={118},
  number={15},
  year={2021},
  publisher={National Acad Sciences}
}

@article{dushoff2021transmission,
  title={Transmission dynamics are crucial to COVID-19 vaccination policy},
  author={Dushoff, Jonathan and Colijn, Caroline and Earn, David JD and Bolker, Benjamin M},
  journal={Proceedings of the National Academy of Sciences},
  volume={118},
  number={29},
  year={2021},
  publisher={National Acad Sciences}
}

@article{davies2021community,
  title={Community factors and excess mortality in first wave of the COVID-19 pandemic in England},
  author={Davies, Bethan and Parkes, Brandon L and Bennett, James and Fecht, Daniela and Blangiardo, Marta and Ezzati, Majid and Elliott, Paul},
  journal={Nature Communications},
  volume={12},
  number={1},
  pages={1--9},
  year={2021},
  publisher={Nature Publishing Group}
}

@article{gozzi2021estimating,
  title={Estimating the effect of social inequalities on the mitigation of COVID-19 across communities in Santiago de Chile},
  author={Gozzi, Nicol{\`o} and Tizzoni, Michele and Chinazzi, Matteo and Ferres, Leo and Vespignani, Alessandro and Perra, Nicola},
  journal={Nature communications},
  volume={12},
  number={1},
  pages={1--9},
  year={2021},
  publisher={Nature Publishing Group}
}

@article{schmidt2021equitable,
  title={Equitable allocation of COVID-19 vaccines in the United States},
  author={Schmidt, Harald and Weintraub, Rebecca and Williams, Michelle A and Miller, Kate and Buttenheim, Alison and Sadecki, Emily and Wu, Helen and Doiphode, Aditi and Nagpal, Neha and Gostin, Lawrence O and others},
  journal={Nature Medicine},
  pages={1--10},
  year={2021},
  publisher={Nature Publishing Group}
}

@article{dolan1998measurement,
  title={The measurement of individual utility and social welfare},
  author={Dolan, Paul},
  journal={Journal of health economics},
  volume={17},
  number={1},
  pages={39--52},
  year={1998},
  publisher={Elsevier}
}

@Misc{delaware-list,
howpublished = {\url{https://www.coastalpoint.com/news/coronavirus/delaware-list-of-essential-and-non-essential-businesses/pdf_b43d669c-6c8f-11ea-bde7-971f96836402.html}},
%note = {Accessed~on~July~18th,~2021},
title = {Delaware List of Essential and Non-Essential Businesses.},
}

@article{glaeser2020jue,
  title={JUE insight: How much does COVID-19 increase with mobility? Evidence from New York and four other US cities},
  author={Glaeser, Edward L and Gorback, Caitlin and Redding, Stephen J},
  journal={Journal of urban economics},
  pages={103292},
  year={2020},
  publisher={Elsevier}
}

@article{dudley2020disparities,
  title={Disparities in age-specific morbidity and mortality from SARS-CoV-2 in China and the Republic of Korea},
  author={Dudley, Joseph P and Lee, Nam Taek},
  journal={Clinical Infectious Diseases},
  volume={71},
  number={15},
  pages={863--865},
  year={2020},
  publisher={Oxford University Press US}
}

@Misc{household_size,
howpublished = {\url{https://www.statista.com/statistics/242189/disitribution-of-households-in-the-us-by-household-size/}},
%note = {Accessed~on~July~30th,~2021},
title = {US Census Bureau. (December 1, 2020). Distribution of households in the United States from 1970 to 2020, by household size.},
}

@article{goldstein2020demographic,
  title={Demographic perspectives on the mortality of COVID-19 and other epidemics},
  author={Goldstein, Joshua R and Lee, Ronald D},
  journal={Proceedings of the National Academy of Sciences},
  volume={117},
  number={36},
  pages={22035--22041},
  year={2020},
  publisher={National Acad Sciences}
}

@book{tzeng2011multiple,
  title={Multiple attribute decision making: methods and applications},
  author={Tzeng, Gwo-Hshiung and Huang, Jih-Jeng},
  year={2011},
  publisher={CRC press}
}

@Misc{jhuinterim,
howpublished = {\url{https://www.centerforhealthsecurity.org/our-work/publications/interim-framework-for-covid-19-vaccine-allocation-and-distribution-in-the-us}},
%note = {Accessed~on~Aug~15th,~2021},
title = {Interim Framework for COVID-19 Vaccine Allocation and Distribution in the United States.},
organization = {The Johns Hopkins Center for Health Security},
year = {2020}
}

@Misc{safegraphweekly,
howpublished = {\url{https://docs.safegraph.com/docs/weekly-patternss}},
%note = {Accessed~on~Aug~17th,~2021},
title = {Safegraph Weekly Patterns.},
}

@article{macnaughton2009untangling,
  title={Untangling equality and non-discrimination to promote the right to health care for all},
  author={MacNaughton, Gillian},
  journal={health and human rights},
  pages={47--63},
  year={2009},
  publisher={JSTOR}
}

@article{emanuel2020ethical,
  title={An ethical framework for global vaccine allocation},
  author={Emanuel, Ezekiel J and Persad, Govind and Kern, Adam and Buchanan, Allen and Fabre, C{\'e}cile and Halliday, Daniel and Heath, Joseph and Herzog, Lisa and Leland, RJ and Lemango, Ephrem T and others},
  journal={Science},
  volume={369},
  number={6509},
  pages={1309--1312},
  year={2020},
  publisher={American Association for the Advancement of Science}
}

@article{ottersen2008distribution,
  title={Distribution matters: equity considerations among health planners in Tanzania},
  author={Ottersen, Trygve and Mbilinyi, Deogratius and M{\ae}stad, Ottar and Norheim, Ole Frithjof},
  journal={Health Policy},
  volume={85},
  number={2},
  pages={218--227},
  year={2008},
  publisher={Elsevier}
}

@article{lai2020effect,
  title={Effect of non-pharmaceutical interventions to contain COVID-19 in China},
  author={Lai, Shengjie and Ruktanonchai, Nick W and Zhou, Liangcai and Prosper, Olivia and Luo, Wei and Floyd, Jessica R and Wesolowski, Amy and Santillana, Mauricio and Zhang, Chi and Du, Xiangjun and others},
  journal={nature},
  volume={585},
  number={7825},
  pages={410--413},
  year={2020},
  publisher={Nature Publishing Group}
}

@article{leclerc1990differential,
  title={Differential mortality: some comparisons between England and Wales, Finland and France, based on inequality measures},
  author={Leclerc, Annette and Lert, France and FABIEN, C{\'E}CILE},
  journal={International journal of epidemiology},
  volume={19},
  number={4},
  pages={1001--1010},
  year={1990},
  publisher={Oxford University Press}
}

@incollection{illsley1987measurement,
%  title={The measurement of inequality in health},
%  author={Illsley, Raymond and Le Grand, Julian},
%  booktitle={Health and economics},
%  pages={12--36},
%  year={1987},
%  publisher={Springer}
%}

@inproceedings{berndt2003measuring,
  title={Measuring healthcare inequities using the Gini index},
  author={Berndt, Donald J and Fisher, John W and Rajendrababu, Rama V and Studnicki, James},
  booktitle={Proceedings of the 36th Annual Hawaii International Conference on System Sciences, 2003.},
  pages={10--pp},
  year={2003},
  organization={IEEE}
}

@Misc{whocoviddashboard,
%howpublished = {\url{https://covid19.who.int/}},
%note = {Accessed~on~Aug~22th,~ 2021},
%title = {WHO Coronavirus (COVID-19) Dashboard},
%author = {World Health Organization}
%}

@Misc{nytdata,
howpublished = {\url{https://github.com/nytimes/covid-19-data}},
%note = {Accessed~on~Feb~6th,~2021},
title = {Coronavirus (Covid-19) Data in the United States.},
author = {The~New~York~Times}
}

@Misc{cbgdef,
howpublished = {\url{https://en.wikipedia.org/wiki/Census_block_group}},
%note = {Accessed~on~Aug~23th,~2021},
title = {Definition of census block group.},
}

@Misc{ew-unitedway,
howpublished = {\url{https://unitedwaynca.org/stories/us-states-essential-workers/}},
%note = {Accessed~on~Aug~23th,~2021},
title = {US States with the most essential workers.},
}

@Misc{worldbanklabor,
howpublished = {\url{https://data.worldbank.org/indicator/SL.TLF.TOTL.IN?locations=US}},
%note = {Accessed~on~Aug~23th,~2021},
title = {Labor force in the United States.},
}

@article{backman2008health,
  title={Health systems and the right to health: an assessment of 194 countries},
  author={Backman, Gunilla and Hunt, Paul and Khosla, Rajat and Jaramillo-Strouss, Camila and Fikre, Belachew Mekuria and Rumble, Caroline and Pevalin, David and P{\'a}ez, David Acurio and Pineda, M{\'o}nica Armijos and Frisancho, Ariel and others},
  journal={The Lancet},
  volume={372},
  number={9655},
  pages={2047--2085},
  year={2008},
  publisher={Elsevier}
}

@article{PILLAY20082005,
title = {Right to health and the Universal Declaration of Human Rights},
journal = {The Lancet},
volume = {372},
number = {9655},
pages = {2005-2006},
year = {2008},
%issn = {0140-6736},
%doi = {https://doi.org/10.1016/S0140-6736(08)61783-3},
%url = {https://www.sciencedirect.com/science/article/pii/S0140673608617833},
author = {Navanethem Pillay}
}

@article{2013B5,
title = {Global Vaccine Action Plan},
journal = {Vaccine},
volume = {31},
pages = {B5-B31},
year = {2013},
note = {Decade of Vaccines},
%issn = {0264-410X},
%doi = {https://doi.org/10.1016/j.vaccine.2013.02.015},
%url = {https://www.sciencedirect.com/science/article/pii/S0264410X13001680}
}

@article{wang2020combating,
  title={Combating COVID-19: health equity matters},
  author={Wang, Zhicheng and Tang, Kun},
  journal={Nature medicine},
  volume={26},
  number={4},
  pages={458--458},
  year={2020},
  publisher={Nature Publishing Group}
}

@article{maxmen2021covid,
  title={COVID boosters for wealthy nations spark outrage.},
  author={Maxmen, Amy},
  journal={Nature},
  year={2021}
}

@article{padma2021covid,
  title={COVID vaccines to reach poorest countries in 2023—Despite recent pledges},
  author={Padma, TV},
  journal={Nature},
  volume={595},
  number={7867},
  pages={342--343},
  year={2021},
  publisher={Nature}
}

@article{sassi2001equity,
  title={Equity versus efficiency: a dilemma for the NHS.},
  author={Sassi, F and Le Grand, J and Titmuss, R},
  journal={BMJ: British Medical Journal: International Edition},
  volume={323},
  number={7316},
  pages={762--763},
  year={2001},
  publisher={BMJ Publishing Group}
}

@article{dixon1987bootstrapping,
  title={Bootstrapping the Gini coefficient of inequality},
  author={Dixon, Philip M and Weiner, Jacob and Mitchell-Olds, Thomas and Woodley, Robert},
  journal={Ecology},
  volume={68},
  number={5},
  pages={1548--1551},
  year={1987},
  publisher={JSTOR}
}

@techreport{bentham1781introduction,
  title={An introduction to the principles of morals and legislation},
  author={Bentham, Jeremy},
  year={1781},
  institution={McMaster University Archive for the History of Economic Thought}
}

@article{buckner2021dynamic,
  title={Dynamic prioritization of COVID-19 vaccines when social distancing is limited for essential workers},
  author={Buckner, Jack H and Chowell, Gerardo and Springborn, Michael R},
  journal={Proceedings of the National Academy of Sciences},
  volume={118},
  number={16},
  year={2021},
  publisher={National Acad Sciences}
}

@Misc{prioritize_eu,
howpublished = {\url{https://www.ecdc.europa.eu/en/news-events/ecdc-releases-vaccination-rollout-strategies-eueea}},
title = {ECDC releases COVID-19 vaccination rollout strategies for EU/EEA.},
year = {2020},
organization={European Centre for Disease Prevention and Control},
%note = {Accessed~on~Sept~15th,~2021},
}

@Misc{prioritize_us,
howpublished = {\url{https://disabilityhealth.jhu.edu/vaccine-2/}},
title = {Vaccine Prioritization Dashboard.},
organization={Johns Hopkins University},
%note = {Accessed~on~Sept~15th,~2021},
}

@Misc{svi2018,
howpublished = {\url{https://www.atsdr.cdc.gov/placeandhealth/svi/data_documentation_download.html}},
title = {CDC/ATSDR Social Vulnerability Index 2018.},
organization={Centers for Disease Control and Prevention and Agency for Toxic Substances and Disease Registry/Geospatial Research, Analysis, and Services Program},
%note = {Accessed~on~Sept~15th,~2021},
}

@book{Golan2021,
author="Golan, Maureen S.
and Trump, Benjamin D.
and Cegan, Jeffrey C.
and Linkov, Igor",
editor="Linkov, Igor
and Keenan, Jesse M.
and Trump, Benjamin D.",
title="The Vaccine Supply Chain: A Call for Resilience Analytics to Support COVID-19 Vaccine Production and Distribution",
bookTitle="COVID-19: Systemic Risk and Resilience",
year="2021",
publisher="Springer International Publishing",
address="Cham",
pages="389--437",
%isbn="978-3-030-71587-8",
%doi="10.1007/978-3-030-71587-8_22",
%url="https://doi.org/10.1007/978-3-030-71587-8_22"
}

@article{jama_equity,
    author = {Jean-Jacques, Muriel and Bauchner, Howard},
    title = "{Vaccine Distribution—Equity Left Behind?}",
    journal = {JAMA},
    volume = {325},
    number = {9},
    pages = {829-830},
    year = {2021},
    month = {03},
    %issn = {0098-7484},
    doi = {10.1001/jama.2021.1205},
    %url = {https://doi.org/10.1001/jama.2021.1205},
    %eprint = {https://jamanetwork.com/journals/jama/articlepdf/2776053/jama\_jeanjacques\_2021\_ed\_210007\_1614619866.67294.pdf},
}

@article{sanchez2020jobs,
  title={Which jobs are most vulnerable to COVID-19? What an analysis of the European Union reveals},
  author={Sanchez, Daniel Garrote and Parra, Nicolas Gomez and Ozden, Caglar and Rijkers, Bob},
  journal={World Bank Research and Policy Briefs},
  number={148384},
  year={2020}
}

@Misc{svi_michigan,
howpublished = {\url{https://www.michigan.gov/documents/coronavirus/Social_Vulnerability_and_COVID-19-v4_715525_7.pdf}},
title = {Social Vulnerability \& COVID-19.},
organization={Michigan Department of Health \& Human Services (MDHHS)},
%note = {Accessed~on~Sept~15th,~2021},
}

@article{persad2009principles,
  title={Principles for allocation of scarce medical interventions},
  author={Persad, Govind and Wertheimer, Alan and Emanuel, Ezekiel J},
  journal={The Lancet},
  volume={373},
  number={9661},
  pages={423--431},
  year={2009},
  publisher={Elsevier}
}

@article{doi:10.1056/NEJMsb2005114,
author = {Emanuel, Ezekiel J. and Persad, Govind and Upshur, Ross and Thome, Beatriz and Parker, Michael and Glickman, Aaron and Zhang, Cathy and Boyle, Connor and Smith, Maxwell and Phillips, James P.},
title = {Fair Allocation of Scarce Medical Resources in the Time of Covid-19},
journal = {New England Journal of Medicine},
volume = {382},
number = {21},
pages = {2049-2055},
year = {2020},
%doi = {10.1056/NEJMsb2005114},
%URL = {https://doi.org/10.1056/NEJMsb2005114},
%eprint = {https://doi.org/10.1056/NEJMsb2005114}
}

@article{bubar2021model,
  title={Model-informed COVID-19 vaccine prioritization strategies by age and serostatus},
  author={Bubar, Kate M and Reinholt, Kyle and Kissler, Stephen M and Lipsitch, Marc and Cobey, Sarah and Grad, Yonatan H and Larremore, Daniel B},
  journal={Science},
  volume={371},
  number={6532},
  pages={916--921},
  year={2021},
  publisher={American Association for the Advancement of Science}
}

@article{fitzpatrick2021optimizing,
  title={Optimizing age-specific vaccination},
  author={Fitzpatrick, Meagan C and Galvani, Alison P},
  journal={Science},
  volume={371},
  number={6532},
  pages={890--891},
  year={2021},
  publisher={American Association for the Advancement of Science}
}

@article{matrajt2021vaccine,
  title={Vaccine optimization for COVID-19: Who to vaccinate first?},
  author={Matrajt, Laura and Eaton, Julia and Leung, Tiffany and Brown, Elizabeth R},
  journal={Science Advances},
  volume={7},
  number={6},
  pages={eabf1374},
  year={2021},
  publisher={American Association for the Advancement of Science}
}

@article{khubchandani2021covid,
  title={COVID-19 vaccination hesitancy in the United States: a rapid national assessment},
  author={Khubchandani, Jagdish and Sharma, Sushil and Price, James H and Wiblishauser, Michael J and Sharma, Manoj and Webb, Fern J},
  journal={Journal of Community Health},
  volume={46},
  number={2},
  pages={270--277},
  year={2021},
  publisher={Springer}
}

@article{dror2020vaccine,
  title={Vaccine hesitancy: the next challenge in the fight against COVID-19},
  author={Dror, Amiel A and Eisenbach, Netanel and Taiber, Shahar and Morozov, Nicole G and Mizrachi, Matti and Zigron, Asaf and Srouji, Samer and Sela, Eyal},
  journal={European journal of epidemiology},
  volume={35},
  number={8},
  pages={775--779},
  year={2020},
  publisher={Springer}
}

@article{loomba2021measuring,
  title={Measuring the impact of COVID-19 vaccine misinformation on vaccination intent in the UK and USA},
  author={Loomba, Sahil and de Figueiredo, Alexandre and Piatek, Simon J and de Graaf, Kristen and Larson, Heidi J},
  journal={Nature human behaviour},
  volume={5},
  number={3},
  pages={337--348},
  year={2021},
  publisher={Nature Publishing Group}
}

@Misc{vac_demo,
%howpublished = {\url{https://covid.cdc.gov/covid-data-tracker/\#\#vaccination-demographic}},
%title = {COVID Data Tracker, Vaccination Demographics.},
%organization={Centers for Disease Control and Prevention, U.S.},
%%note = {Accessed~on~Sept~15th,~2021},
%}

\section*{Acknowledgments}
\textbf{Funding:}
This work was supported in part by The National Key Research and Development Program of China under grant 2020AAA0106000 and the National Natural Science Foundation of China under U1936217.\\
\textbf{Author Contributions:}
Fengli Xu, Pan Hui, Yong Li, and James Evans jointly launched this research and provided the research outline. Lin Chen, Fengli Xu and Yong Li designed the research methods. Lin Chen performed the experiments and prepared the figures. Fengli Xu, Zhenyu Han, Kun Tang, Yong Li, Pan Hui and James Evans provided critical revisions. All authors jointly analyzed the results and participated in the writing of the manuscript.\\
\textbf{Competing interests:}
The authors declare no competing interests.\\
\textbf{Data and materials availability:}
All data and code used in this research are publicly available.

\section*{Supplementary materials}
Materials and Methods\\
Figs S1 to S17\\
Tables S1 to S14\\
References (58-69)


\clearpage


\begin{figure}[h]
    \centering
    \includegraphics[width=0.95\textwidth]{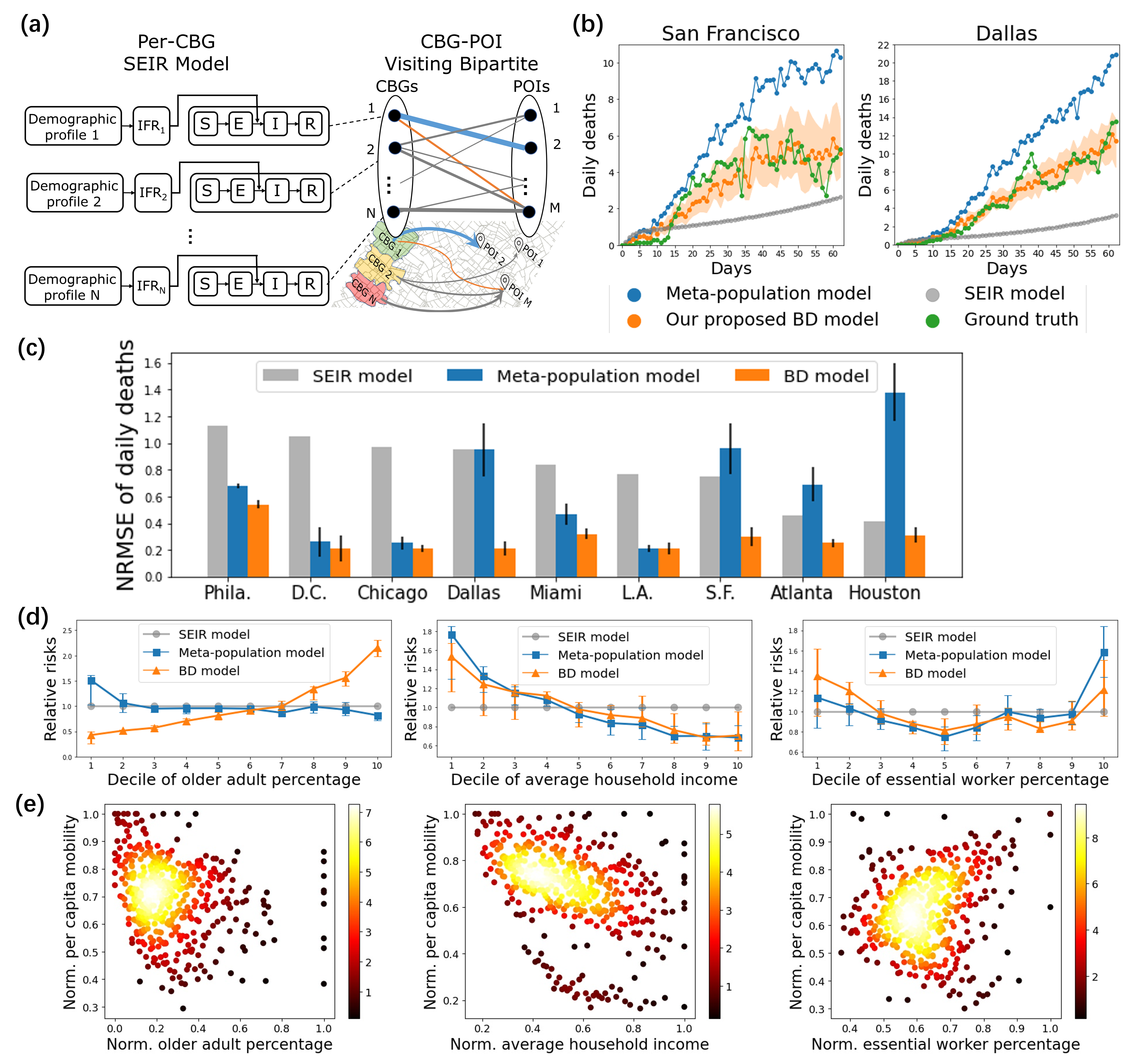}
    \caption{\textbf{\underline{B}ehavior and \underline{D}emography informed epidemic modelling.}
    \textbf{(a)} Overview of our \underline{B}ehavior and \underline{D}emography informed epidemic model (BD model).
    \textbf{(b)} Capability in fitting to representative curves of daily deaths.  
    Whether daily deaths grow sub-linearly or almost linearly, 
    our BD model (orange line) fits more accurately to the ground truth (green line), compared to the SEIR model (grey line) and the meta-population model (blue line). 
    The shaded regions show results of parameter sets that achieve an RMSE within 150\% of the best result. 
    \textbf{(c)} 
    Our BD model (orange bars) reduces the normalized RMSE in daily death prediction by up to $80\%$ and $77\%$ compared with the SEIR model (grey bars) and the meta-population model (blue bars), respectively.
    \textbf{(d)} Predicted uneven fatality rates among communities with different demographic features.
    Our BD model (orange dots) successfully captures the high mortality risks faced by communities with high percentages of older adults (left panel), low household income (middle panel), and high percentages of essential workers (right panel), while two baseline models fail.
    \textbf{(e)} Joint distribution of demographic features and mobility.
    The percentage of older adults and average household income negatively correlate with per capita mobility ($r=-0.29$, $r=-0.45$, respectively).
    The percentage of essential workers positively correlates with per capita mobility ($r=0.39$).}
    \label{fig1}
\end{figure}

\newpage
\begin{figure}[h]
    \centering
    \includegraphics[width=1\textwidth]{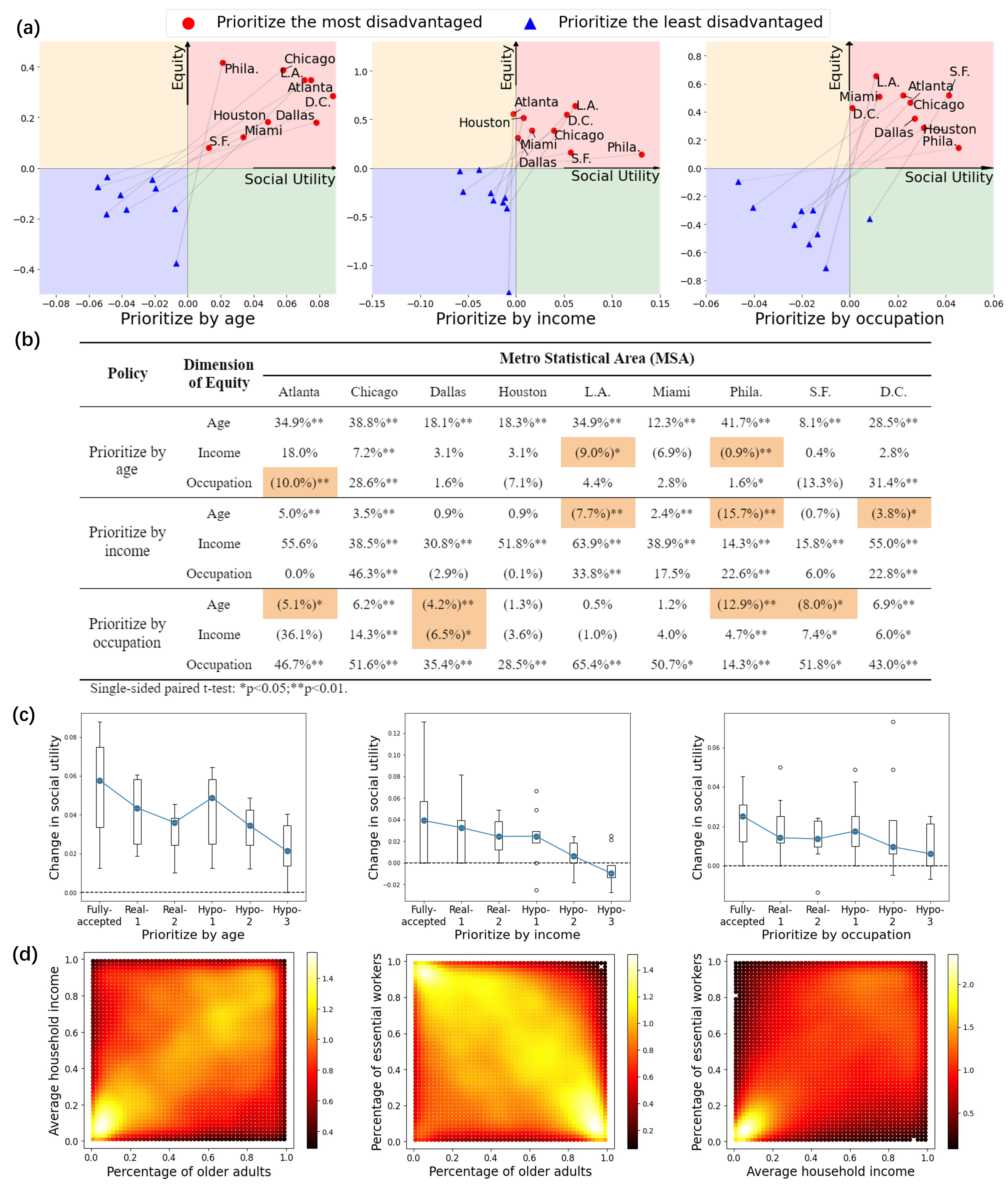}
    \caption{\textbf{Social utility and equity under different vaccine distribution strategies.}}
    \label{fig2}
\end{figure}

\clearpage
\begin{spacing}{1.0}
\noindent\textbf{Figure 2: Social utility and equity of vaccine distribution strategies that prioritize disadvantaged communities under a single demographic dimension.}
\textbf{(a)} Changes in social utility and equity, compared to the \textit{Homogeneous} baseline.
The red/blue points respectively represent the strategies prioritizing the most/least disadvantaged communities.
The $1^{st}$ to $4^{th}$ quadrant respectively represents: (i) simultaneously improving utility and equity, (ii) improving equity but damaging utility, (iii) improving utility but damaging equity, and (iv) simultaneously damaging utility and equity.
\textbf{(b)} Changes in three dimensions of equity, compared to the \textit{Homogeneous} baseline.
Highlighted grids indicate degradation in the corresponding dimension of equity.
\textbf{(c)} 
Social utility under different scenarios of vaccine hesitancy.
When vaccine hesitancy in low-income communities is stronger, the benefit on social utility brought by prioritizing disadvantaged communities diminishes.
In extreme scenarios, the benefit of prioritizing disadvantaged communities characterized by income is completely erased, making it inferior to the baseline. 
\textbf{(d)} 
Joint probability distribution of demographic features, where brighter colors indicate larger probability density.
The correlations between (i) the percentage of older adults and the average household income, (ii) the percentage of older adults and the percentage of essential workers, (iii) the average household income and the percentage of essential workers are (i) $+0.14$, (ii) $-0.2$, and (iii) $+0.28$, respectively, displaying the mismatch among disadvantaged communities in different demographic dimensions.
\end{spacing}

\newpage
\begin{figure}[h]
    \centering
    \includegraphics[width=1\textwidth]{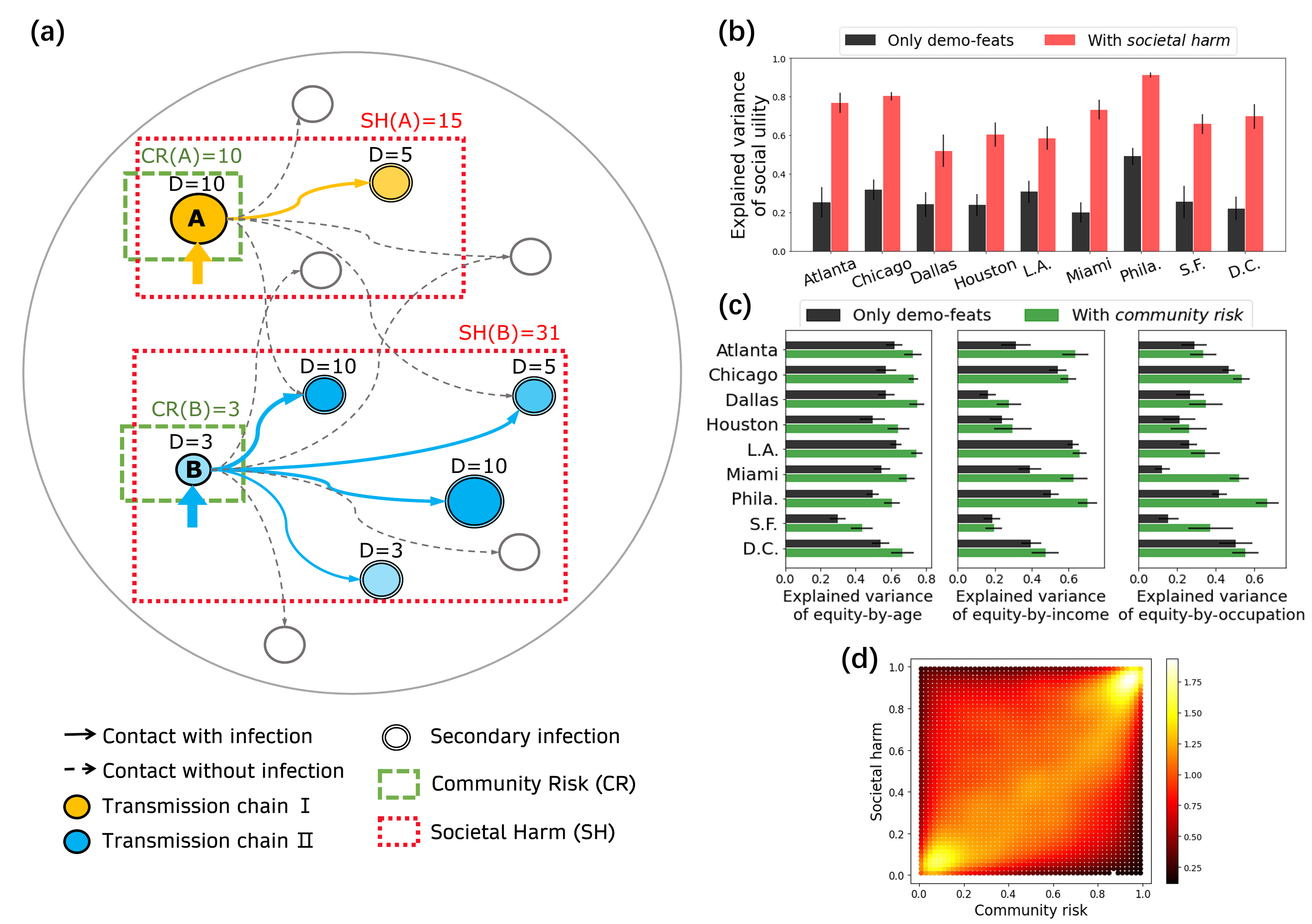}
    \caption{\textbf{Design and justification of \textit{community risk} and \textit{societal harm}.}
    \textbf{(a)} Illustration of \textit{community risk} (\textit{CR}) and \textit{societal harm} (\textit{SH}). 
    Each node represents a community, whose size reflects the community's vulnerability, and color tint reflects the number of deaths in the community, quantified by the value of $D$. 
    Each edge represents the inter-community mobility connections, whose thickness reflects the mobility intensity.
    For each community, \textit{CR} equals its own mortality risk (green boxes), and \textit{SH} equals the sum of its own mortality risk and the mortality risk it potentially presents to others (red boxes).
    As two representative cases, community $A$ of transmission chain $\uppercase\expandafter{\romannumeral1}$ has large \textit{CR}  but small \textit{SH}, while community $B$ of transmission chain $\uppercase\expandafter{\romannumeral2}$ has small \textit{CR}  but large \textit{SH}.
    \textbf{(b)} OLS regression of changes in social utility with/without \textit{societal harm}.
    Regressions with only demographic features explain on average $28.0\%$ of the variance, measured by adjusted $R^2$ (black bars).
    The incorporation of \textit{societal harm} raises the explained variance to an average of $62.0\%$ (red bars), greatly improving the goodness-of-fit of the regression model.
    \textbf{(c)} OLS regression of changes in equity with/without \textit{community risk}.
    Regressions with only demographic features explain on average $53.0\%$, $37.3\%$ and $29.9\%$ of the variance, respectively (black bars).
    The incorporation of \textit{community risk} raises the explained variance to an average of $66.4\%$, $49.6\%$ and $43.8\%$, respectively (green bars), greatly improving the goodness-of-fit of the regression model.
    \textbf{(d)} Joint probability distribution of \textit{community risk} and \textit{societal harm}, where brighter colors indicate larger probability density.
    There is a non-negligible positive correlation ($r=0.29$) between \textit{community risk} and \textit{societal harm}.
    }
    \label{fig3}
\end{figure}

\newpage
\begin{figure}[h]
    \centering
    \includegraphics[width=0.98\textwidth]{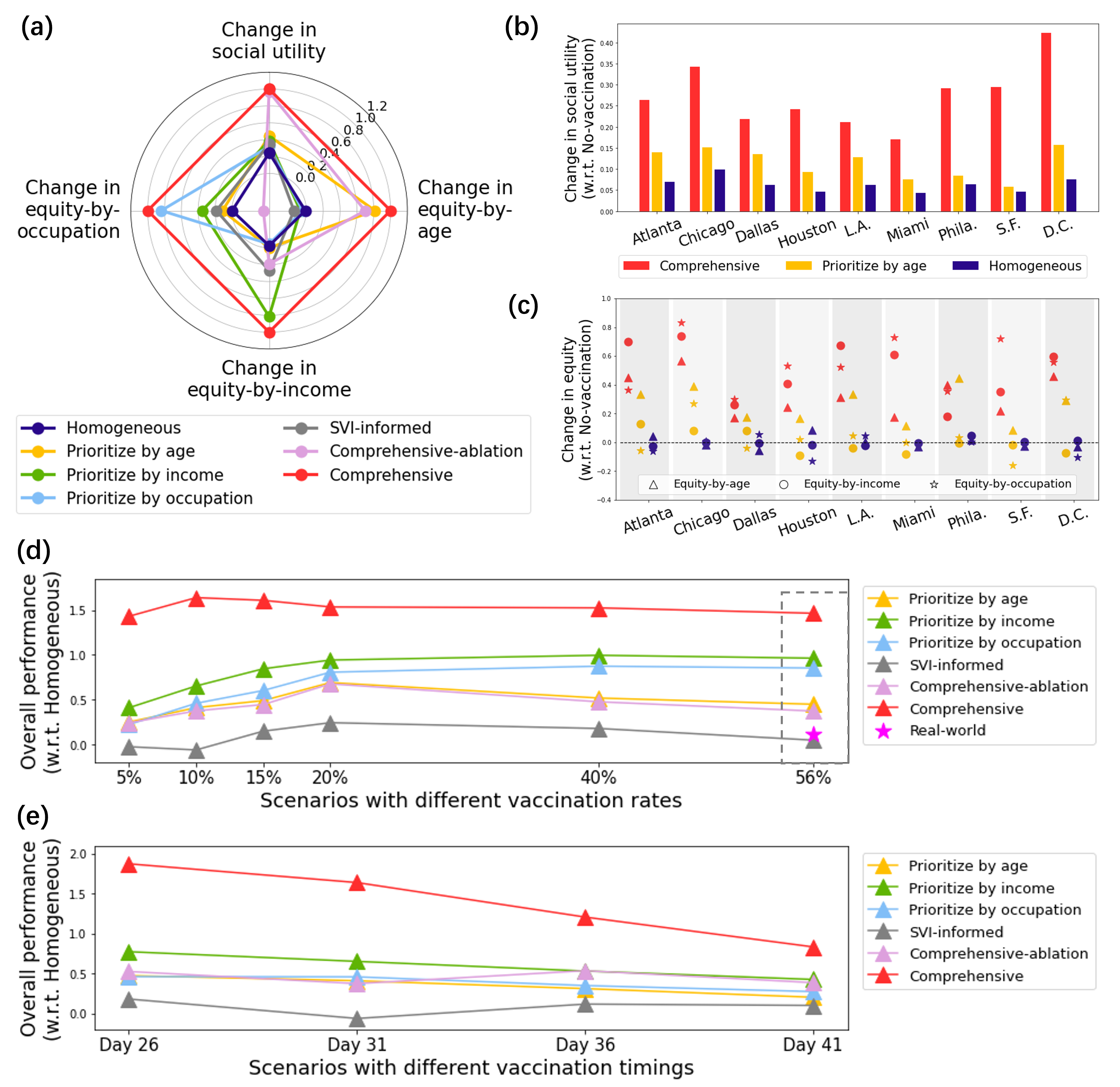}
    \caption{\textbf{Performance of the \textit{Comprehensive} distribution strategy under various vaccination rates and timings.}
    \textbf{(a)} Changes in social utility and three dimensions of equity under 7 vaccine distribution strategies. 
    The designed \textit{Comprehensive} strategy (red line) encompasses all the other strategies in four metrics, indicating its well-rounded effectiveness.
    In contrast, \textit{SVI-informed} (grey line) and \textit{Comprehensive-ablation} (violet line) result in degradation in certain dimension of equity. 
    \textbf{(b)} Changes in social utility in each MSA.
    \textbf{(c)} Changes in equity by age, income and occupation in each MSA. 
    \textbf{(d)} Overall performance of strategies under different vaccination rates. 
    Overall performance is the sum of relative improvement in social utility and three dimensions of equity compared to the \textit{Homogeneous} baseline.
    The star shows the overall performance if a vaccine is distributed proportionally to its real-world distribution, with a vaccination rate of $56\%$ that is close to the current rate in the U.S..
    \textbf{(e)} Overall performance of strategies under different vaccination timings.
    }
    \label{fig4}
\end{figure}

\end{document}